\documentclass[fleqn,usenatbib]{mnras}

\usepackage{newtxtext,newtxmath}

\usepackage[T1]{fontenc}

\DeclareRobustCommand{\VAN}[3]{#2}
\let\VANthebibliography\thebibliography
\def\thebibliography{\DeclareRobustCommand{\VAN}[3]{##3}\VANthebibliography}

\usepackage{graphicx}	
\usepackage{amsmath}	
\usepackage{subcaption}

\newcommand{\HI}{H{\sc i}}	

\newcommand{\Msol}{\rm{M}_{\odot}}

\defcitealias{Obreschkow2016}{O16}
\defcitealias{Hardwick2022a}{Paper 1}
\defcitealias{Hardwick2022b}{Paper 2}

\title[JMG plane for {\sc EAGLE} and IllustrisTNG]{Exploring the Angular Momentum -- Atomic Gas Content Connection with {\sc EAGLE} and IllustrisTNG}

\author[J. A. Hardwick et al.]{Jennifer A. Hardwick,$^{1,2}$\thanks{E-mail: \href{mailto:jennifer.hardwick@icrar.org}{jennifer.hardwick@icrar.org}}
Luca Cortese,$^{1,2}$
Danail Obreschkow,$^{1,2}$
Claudia Lagos,$^{1,2}$
Adam R. H. Stevens,$^{1,2}$
\newauthor
Barbara Catinella,$^{1,2}$
and Lilian Garratt-Smithson$^{1,2}$
\\
$^{1}$International Centre for Radio Astronomy Research (ICRAR), University of Western Australia, Crawley, WA 6009, Australia\\
$^{2}$Australian Research Council, Centre of Excellence for All Sky Astrophysics in 3 Dimensions (ASTRO 3D), Australia\\
}

\date{Accepted 2023 September 14. Received 2023 August 28; in original form 2023 July 05}

\pubyear{2023}

\begin{document}
\label{firstpage}
\pagerange{\pageref{firstpage}--\pageref{lastpage}}
\maketitle


\begin{abstract}

We use the {\sc EAGLE} (Evolution and Assembly of GaLaxies and their Environments) and IllustrisTNG (The Next Generation) cosmological simulations to investigate the properties of the baryonic specific angular momentum (j), baryonic mass (M) and atomic gas fraction ($f_{\rm{atm}}$) plane for nearby galaxies.
We find EAGLE and TNG to be in excellent agreement with each other. These simulations are also consistent with the results obtained with xGASS (eXtended GALEX Arecibo SDSS Survey) for gas fractions greater than 0.01. 
This implies that the disagreements previously identified between xGASS and predictions from simple analytical disc stability arguments also holds true for {\sc EAGLE} and {\sc TNG}. 
For lower gas fraction (the regime currently unconstrained by observations), both simulations deviate from the plane but still maintain good agreement with each other. 
Despite the challenges posed by resolution limits at low gas fractions, our findings suggest a potential disconnect between angular momentum and gas fraction in the gas-poor regime, implying that not all gas-poor galaxies have low specific angular momentum.

\end{abstract}




\section{Introduction}

Angular momentum is a key property of galaxies, as it is linked to their formation and evolutionary history.
It is now known that the stellar angular momentum scales with mass (the so-called Fall relation; \citealt{Fall1983, Fall2013, Sweet2018, Posti2018a, Posti2018b, Lapi2018, Stone2021, Du2022, DiTeodoro2023}), and that the scatter in the relation is correlated with morphology and stellar structure \citep[e.g.,][]{Fall1983, Romanowsky2012, Cortese2016, Pulsoni2023}.
However, it is still unclear whether stellar structure is the primary physical driver of the scatter in the Fall relation or simply a proxy of the overall accretion history of galaxies.
Indeed, from a theoretical point of view, the growth of angular momentum of discs should be tightly connected to their ability to accrete gas \citep[e.g.,][]{Mo1998, Boissier2000}, potentially implying a more fundamental role of cold gas in driving the scatter of the relation. 
This is also consistent with theoretical work focused on the link between gas content and disc stability \citep[e.g.,][]{Obreschkow2016, Stevens2018, Romeo2020}.
The well-motivated theory behind the connection of a galaxy's gas content and specific angular momentum has been empirically tested with modest observational samples, but both the sample size and manner in which those data are analysed can be improved.

The advent of new datasets allows more detailed investigations of the role of gas content in the build-up of angular momentum in galaxies.
In particular, \citet[][hereafter \citetalias{Hardwick2022a}]{Hardwick2022a} used the extended GALEX Arecibo SDSS Survey (xGASS, \citealt{Catinella2010, Catinella2018}) sample, a deep \HI\ survey which is representative of the local Universe, to study the stellar Fall relation. 
We found that the most strongly correlated parameter with the scatter of the Fall relation is \HI\ gas fraction, not bulge-to-total ratio, particularly for low stellar masses and when isolating the disc component of galaxies. 
In \citet[][hereafter \citetalias{Hardwick2022b}]{Hardwick2022b} we expanded this work and investigated the connection between a galaxy's baryonic angular momentum, baryonic mass and atomic gas fraction, and found that a tight plane exists between these three parameters. 
However, this plane deviates from predictions from simple analytical models of disc stability \citep[e.g.,][]{Obreschkow2016, Romeo2020}, which predict a steeper slope and, most importantly, a stronger dependence on gas fraction. 
A similar result was also found by \citet{Pina2021, Pina2021b} studying a sample of $\sim$100 disk galaxies with resolved HI rotation curves, suggesting that more detailed modelling is needed to fully unveil the physical connection between mass, angular momentum and cold gas content.

The natural next step is to extend such a comparison between xGASS and theoretical models to cosmological hydrodynamical simulations, which are not limited by simplifying assumptions of the models above. 
While recent years have seen a dramatic increase in the number of studies focused on the origin and drivers of the scatter of the Fall relation \citep{Teklu2015, Genel2015, Lagos2017, Zoldan2018, Wang2019, Marshall2019, Stevens2019, Elson2023}.
In particular, qualitative results from \cite{Rodriguez-Gomez2022} suggests that observational trends from \citetalias{Hardwick2022a} are reproduced in IllustrisTNG.
However, no work to date has explicitly investigated the baryonic specific angular momentum ($j_{\rm{bar}}$) -- baryonic mass ($M_{\rm{bar}}$) -- atomic gas fraction ($f_{\rm{atm}}$) plane (hereafter, the JMG plane), quantified its slope and scatter, and carried out a detailed comparison with observations. 
This is critical not only to obtain some insights into the physics driving this relation, but also to investigate how universal this JMG plane really is. 

Despite the xGASS sample showing a greater diversity of galaxies than surveys of just disk galaxies, the sub-sample for which angular momentum can be estimated from \HI\ line widths is still biased towards galaxies that have \HI\ contents above the detection limit of the survey. 
Additionally, although xGASS is a large sample in the context of similar observational samples, the number of galaxies is quite small when compared to large simulation volumes.
Therefore, in this work, we wish to further test the robustness of this JMG plane with even greater statistics and galaxy diversity by using large hydrodynamical cosmological simulations.
First, we aim to test how well the observations and simulations agree in this parameter space. 
Then we can use the simulation data to better explore the physical connection between these three parameters.
As these simulations are not limited to observational constraints, they allow us to probe  
down to low gas fractions and weaker rotational support than is possible with observational surveys.

This paper is set out as follows. 
We start by describing the archival observational and simulation data used in this work from xGASS, {\sc EAGLE} (Evolution and Assembly of GaLaxies and their Environments; \citealt{Schaye2015, Crain2015}) and IllustrisTNG (Illustris The Next Generation, hereafter, {\sc TNG}; \citealt{Pillepich2018b, Nelson2018}) in section \ref{section: data}.
We then explain how we create mock detection samples for the simulations in section \ref{section: mock obs sample}.
The results of comparing {\sc EAGLE} and {\sc TNG} to xGASS in the $M_{\rm{bar}}$ - $j_{\rm{bar}}$ - $f_{\rm{atm}}$ parameter space is presented in section \ref{section: results}. We then discuss the implications of these results in section \ref{section: discussion} and conclude in section \ref{section: conclusion}.

\section{Data description} \label{section: data}

\subsection{xGASS}
Our observational dataset comes from xGASS \citep{Catinella2010,Catinella2018}, which includes galaxies in the stellar mass range of $10^{9}$ to $10^{11.5}\Msol$ across the redshift range $0.01<z<0.05$ and was selected from SDSS DR6 \citep{Adelman-McCarthy2008}. 
Galaxies were observed with the Arecibo telescope until detected in \HI\ or until a gas-fraction limit of 2--10 per cent was reached. 
The sample, in particular at high stellar masses, was selected to have a nearly flat stellar mass distribution. 
Overall, xGASS represents arguably the best representative sample of integrated gas properties in the local Universe.
In this work, we use the same sub-sample of 564 xGASS galaxies that were used in \citetalias{Hardwick2022a} and \citetalias{Hardwick2022b}.
Briefly, this includes only galaxies detected in HI and for which we could accurately determine kinematics; i.e., an inclination greater than 30 degrees and not affected by confusion within the radio beam.
We use the stellar and baryonic specific AM that was calculated in \citetalias{Hardwick2022a} and \citetalias{Hardwick2022b} respectively, which are publicly available.\footnote{\url{xgass.icrar.org}}
These were estimated by combining stellar mass surface density profiles (within a $10 R_{e}$ aperture) with \HI\ widths.
AM was calculated within a large aperture to ensure convergence for galaxies with high S\'{e}rsic indices \citep{Sersic1963}.

\subsection{{\sc EAGLE} and {\sc TNG}}

In this work, we compare our observational results to simulation data from both the {\sc EAGLE} \citep{Schaye2015, Crain2015} cosmological hydrodynamical simulation and {\sc TNG} \citep{Nelson2019,Pillepich2018b, Springel2018, Nelson2018, Naiman2018, Marinacci2018} magnetohydrodynamical cosmological simulation. 
For comparisons to {\sc EAGLE} we use the "reference" {\sc EAGLE} model and AM values determined in \cite{Lagos2017}, and for {\sc TNG} we use the quantities obtained by \cite{Stevens2019}.

For consistency, we use the 100 comoving Mpc simulation box for both {\sc EAGLE} and {\sc TNG}, as they have a comparable number of dark matter particles ($1504^3$ and $1820^3$, respectively).
{\sc EAGLE} is simulated with smoothed particle hydrodynamics using the {\sc gadget}-3 code \citep{Springel2005,Springel2008}, while {\sc TNG} has discretised gas elements within a moving Voronoi mesh implemented using the {\sc arepo} code \citep{Springel2010}.

There are various advantages of simultaneously comparing both works to xGASS. 
First, while both simulations include subgrid models that calculate feedback (from stars and accreting black holes), gas cooling, star formation and black hole growth, the details of the modelling are dramatically different with, for example, the AGN feedback models have significant differences in their physical bases as well as numerical implementations (see \citealt{Schaye2015, Crain2015, Pillepich2018b, Nelson2018} for further details).
Second, the way the subgrid models are calibrated is different. 
{\sc EAGLE} calibrates its subgrid models on two scaling relations; the $z = 0.1$ galaxy stellar mass function and the size--mass relation of disc galaxies. 
In contrast, the primary scaling relations used to calibrate {\sc TNG} subgrid models are the cosmic SFR density history, $z = 0$ galaxy stellar mass function and the $z = 0$ stellar--halo mass relation, with additional scaling relations used as secondary constraints (the black hole--bulge mass relation, the gas fraction of haloes within $R_{500c}$ and the stellar size--mass relation, all at $z = 0$).
Third, the way \cite{Lagos2017} and \cite{Stevens2019} calculated AM is different, as we describe below.

Both \cite{Lagos2017} and \cite{Stevens2019} calculated AM and associated quantities, for the same selection of galaxies; $z = 0$ and $M_{\star} > 10^9 \rm{M}_{\odot}$, (10 803 galaxies for {\sc EAGLE} and 20 876 for {\sc TNG}).
{\sc EAGLE} and {\sc TNG} do not model atomic gas directly; instead, they determine it in post-processing.
We use the \HI\ gas mass from \cite{Lagos2015} and \cite{Stevens2019} that were calculated using the \cite{K13} theoretical model. 
This model determines molecular hydrogen fractions as a function of the total column density of neutral hydrogen, metallicity and the density of the stellar disc, and then uses this to infer atomic hydrogen content (for more details of this model see \citealt{K13}).
$f_{\rm{atm}}$ is then defined to be $1.35 \ M_{\rm{HI}} / M_{\rm{bar}}$, the same as our definition for xGASS (the factor of 1.35 adds the approximate contribution from Helium).
We calculate the baryonic AM as the sum of the AM from stars, \HI\ and $H_{2}$.
In both \cite{Lagos2017} and \cite{Stevens2019} AM is calculated within apertures to be comparable to observations.
From \cite{Lagos2017}, we use the AM calculated within $5R_{e}$ (half-mass radius of stars).
In \cite{Stevens2019} they calculate AM within what they define as the "BaryMP" radius, which is the radius where the gradient of the cumulative baryonic mass profile converges \citep{Stevens2014}. 
This BaryMP radius is $6.9 R_{e}$ on average.

In \citetalias{Hardwick2022a}, for xGASS, we calculated AM within an aperture of $10R_{e}$ to ensure all of our galaxies had their AM converged. 
As $R_{\rm{BaryMP}}$ varies for each galaxy to a radius where the baryonic mass converges, the AM values determined for {\sc TNG} will also likely be converged and comparable with xGASS.
Additionally, if the equations of AM are solved analytically with a single S\'{e}rsic profile \citep{Sersic1963}, a galaxy with a S\'{e}rsic index less than 2 will have its AM converge by $5R_{e}$ (\citetalias{Hardwick2022a}).
Therefore, for the majority of galaxies, the AM determined within $5R_{e}$ or $10R_{e}$ will be comparable and we assume that the values determined in \cite{Lagos2017} are appropriate to compare with our xGASS values.

As we will show in section \ref{section: results}, despite these differences, the agreement between the two simulations and xGASS is striking, highlighting how none of the differences in the way the key parameters investigated here significantly affect our analysis.
To reiterate, these results would remain unchanged even if {\sc EAGLE} AM were calculated in the same way as {\sc TNG}. 
This is because the mock detection samples are dominated by galaxies with a significant disc component, for which AM has already converged to the total value at $5 R_{e}$ (see also \citetalias{Hardwick2022a}).

\section{Mock detection sample} \label{section: mock obs sample}

\begin{figure}
    \centering
    \includegraphics[width=0.5\textwidth]{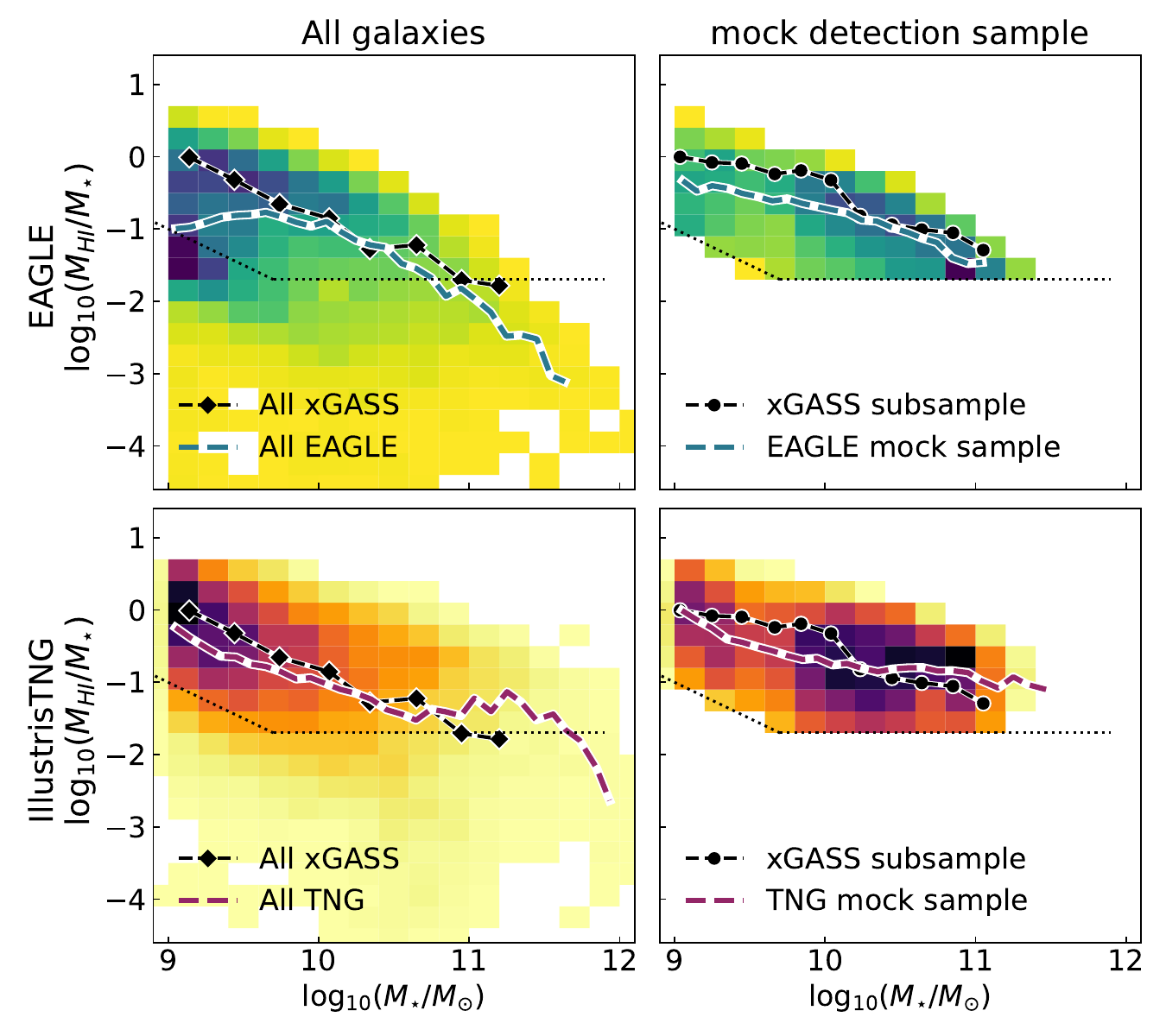}
    \caption{HI gas fraction vs. stellar mass relation, for {\sc EAGLE} (top row) and {\sc TNG} (bottom row). The median relation from the full xGASS sample is shown as black diamonds for comparison (as published in \citealt{Catinella2018}), with the xGASS detection limit shown as dotted black lines. The left column shows all the galaxies in {\sc EAGLE} and {\sc TNG} in the background 2D histogram, with the median in bins of 0.1 dex overlaid as dashed blue/ magenta lines. In the right column, both the {\sc EAGLE} and {\sc TNG} mock detection samples are shown (see text for this selection). Dark colours indicate regions of high density, and vice-versa. The xGASS sub-sample from \citet{Hardwick2022a, Hardwick2022b} is shown as black points. The 2D histogram colours of the left and right columns cannot be compared directly, as the right has been weighted to recover the xGASS stellar distribution, while the left has not.}
    \label{fig: HIfraction-Mass}
\end{figure}

\begin{figure}
    \centering
    \includegraphics[width=0.5\textwidth]{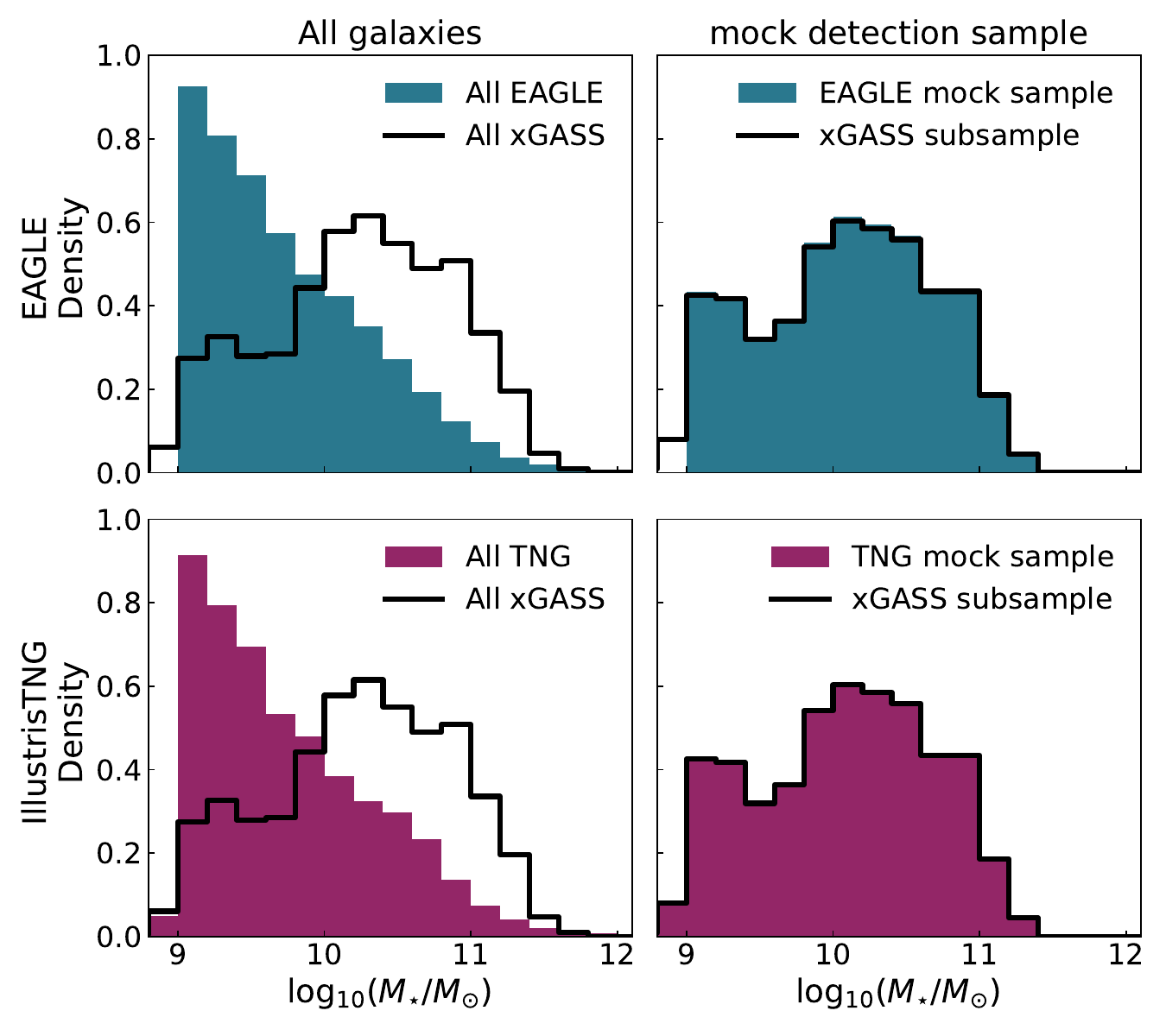}
    \caption{The stellar mass distribution of {\sc EAGLE} (top row, blue) and {\sc TNG} (bottom row, magenta) in comparison to xGASS (black). The left column shows all galaxies in {\sc EAGLE} or {\sc TNG} and compares it to the full xGASS sample (N = 1177). In the right column, we show the xGASS sub-sample used in \citetalias{Hardwick2022a} and \citetalias{Hardwick2022b} (N = 564), compared to {\sc EAGLE} and {\sc TNG} with a HI detection cut applied and a stellar mass weighting to recover the xGASS sub-sample distribution.}
    \label{fig: StellarMassDist}
\end{figure}

The left column of Fig. \ref{fig: HIfraction-Mass} shows the \HI\ gas fraction of all galaxies in {\sc EAGLE} (top row) and {\sc TNG} (bottom row).
The medians for each simulation are represented by blue and magenta dashed lines (for {\sc EAGLE} and {\sc TNG}, respectively).
These can be compared to the weighted median of xGASS in black (these are taken from \citealt{Catinella2018} and in this figure are weighted to recover a volume-limited sample and include \HI\ non-detections).
Both {\sc EAGLE} and {\sc TNG} agree with xGASS for intermediate stellar masses ($\sim 10^{10} M_{\odot}$ to $10^{11} M_{\odot}$), but {\sc EAGLE} galaxies at low stellar masses are more gas--poor than xGASS, while high--stellar--mass {\sc TNG} galaxies are slightly more gas--rich than xGASS (consistent with what was found by \citealt{Dave2020}).
However, it should be noted that the xGASS sample selection is very different from that of the simulations.
The xGASS sample was deliberately selected from SDSS (Sloan Digital Sky Survey, \citealt{Abazajian2009}) to over-sample high stellar masses, which is different to the volume-limited samples of both {\sc EAGLE} and {\sc TNG}. 
This can be seen in the stellar mass distributions shown in the left column of Fig. \ref{fig: StellarMassDist}, where the full xGASS sample is shown as the hollow black histograms, while {\sc EAGLE} and {\sc TNG} are blue and magenta respectively.

To carry out a more fair comparison between xGASS, {\sc EAGLE} and {\sc TNG}, we extract from the simulations a sample that has the same gas fraction limit and stellar mass distribution as xGASS. This can be seen in the right column of Fig. \ref{fig: HIfraction-Mass} and \ref{fig: StellarMassDist}.
This reduces our {\sc EAGLE} sample from 10,803 to 7,037 galaxies and our {\sc TNG} sample from 20,876 to 14,919 galaxies, (although it should be noted that 26 {\sc EAGLE} galaxies and 3,201 {\sc TNG} galaxies had no mass in \HI, so are not shown in the left panel of Fig. \ref{fig: HIfraction-Mass}).
Once the HI detection cut is applied, unsurprisingly the agreement between xGASS and the simulations improves.
In the right column of Fig. \ref{fig: HIfraction-Mass} both samples follow approximately the same distribution with differences now reduced to 0.4 dex or less, primarily at low stellar masses.
The right column of Fig. \ref{fig: StellarMassDist}, {\sc EAGLE} and {\sc TNG} now overlap with the xGASS sub-sample distribution by construction.

Throughout this paper when comparing xGASS to {\sc EAGLE}/ {\sc TNG}, we will show both the full sample and the "mock detection sample" (\HI\ detection cut and stellar mass weighting). 
This allows us to distinguish between physical differences and selection effects.

\section{Results} \label{section: results}

\subsection{Fall Relation}

\begin{figure*}
    \centering
    \begin{subfigure}{0.48\textwidth}
        \includegraphics[width=\textwidth]{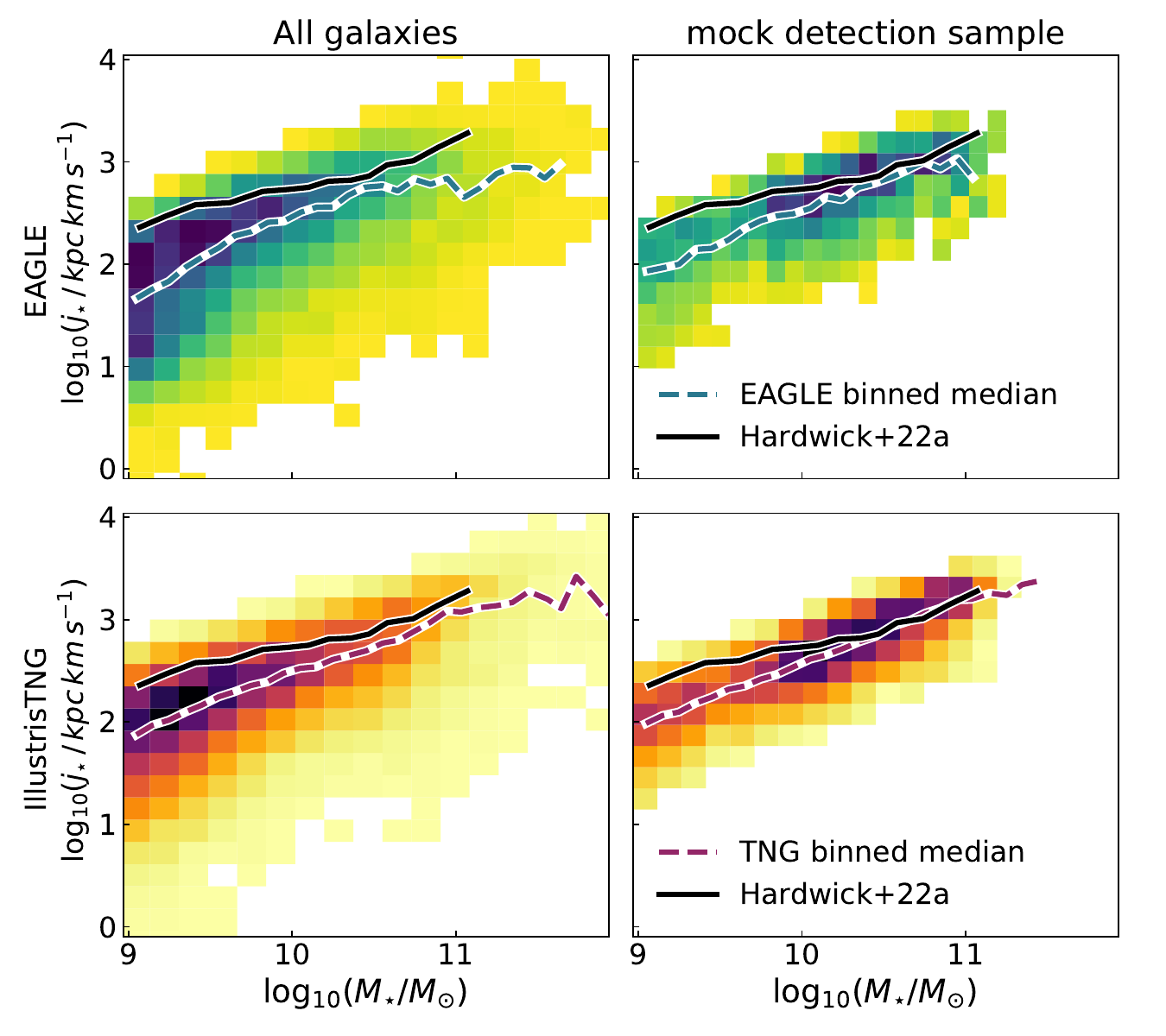}
        \caption{stellar Fall relation}
        \label{fig: StellarFall}
    \end{subfigure}
    \begin{subfigure}{0.48\textwidth}
        \includegraphics[width=\textwidth]{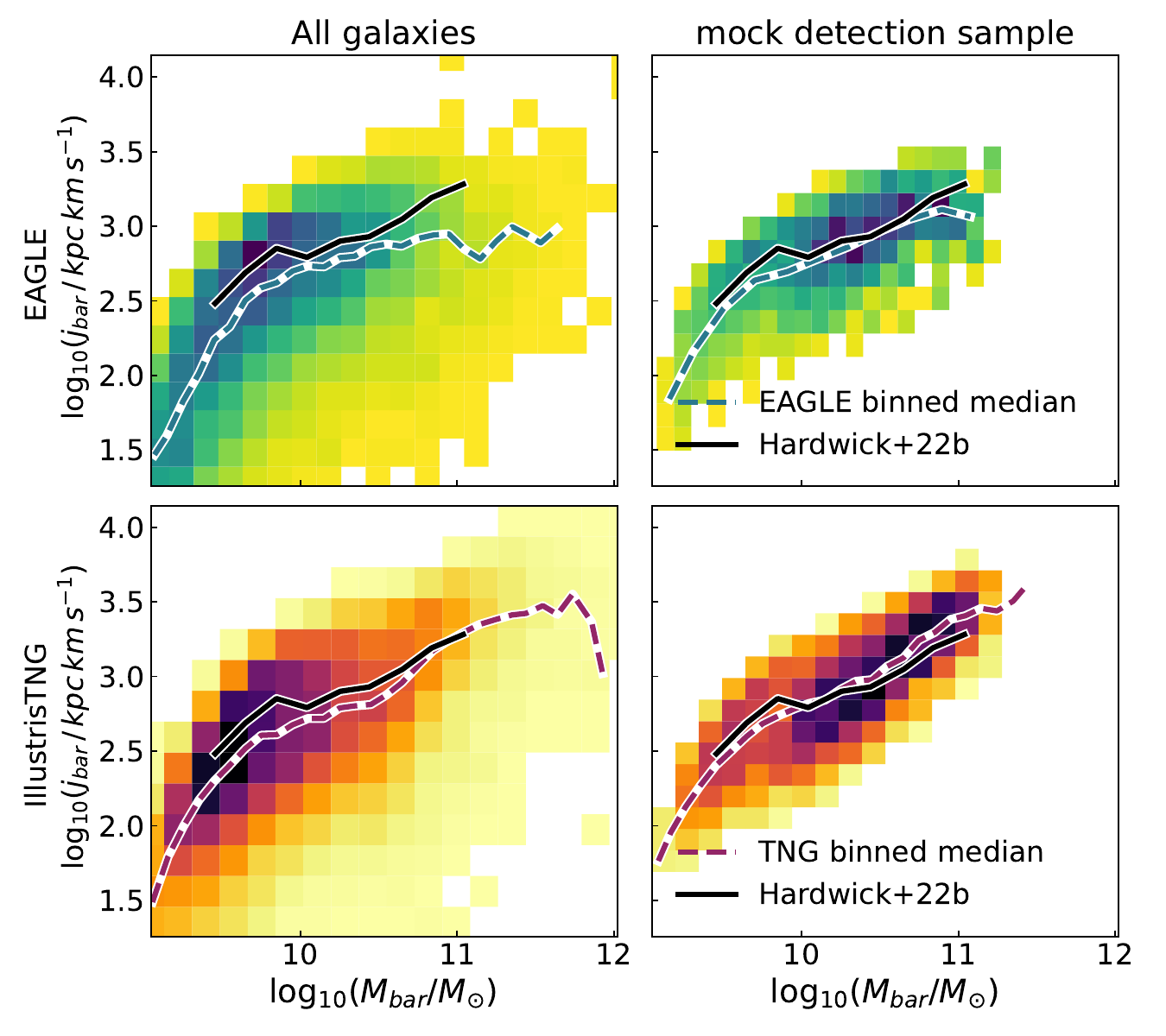}
        \caption{baryonic Fall relation}
        \label{fig: BaryonicFall}
    \end{subfigure}
    \caption{Fall relations of stars (a) and baryons (b). The top row shows the {\sc EAGLE} galaxies, and the bottom row shows the {\sc TNG} galaxies. The left column of each panel is all galaxies in that simulation (above $10^{9} M_{\star} [\rm{M}_{\odot}]$) and the right column is the mock detection sample. In the 2D histograms, dark colours show regions of high density (or in the right column, high weighted density), and vice-versa. Note, the exact colours of the left and right columns can not be compared, as they are on different scales (due to the right being weighted). Dashed coloured lines are medians in bins of width 0.1 dex. In black overlaid are the medians of xGASS galaxies.}
    \label{fig: FallRelations}
\end{figure*}

We first compare {\sc TNG} and {\sc EAGLE} to xGASS observations in the mass--specific AM relation (Fall relation; \citealt{Fall1983}). 
In Fig. \ref{fig: FallRelations}, we show the stellar Fall relations (Panel a) and baryonic Fall relations (Panel b) for both simulations.
The left columns show all of the simulated galaxies without a selection cut, while the right columns are only the galaxies in the simulation above the xGASS detection limit (mock detection sample).
Each axis shows the simulation running median as the coloured dashed line, and a comparison to the xGASS observational sample in black.

We present both columns, to show the effect the sample selection has in this parameter space.
In both the stellar and baryonic cases, the mock detection sample has fewer low AM galaxies than the full samples for both simulations.
This results in a tighter distribution of galaxies around the median for the mock detection sample.
The median AM for low stellar mass galaxies is also higher for the mock detection sample than for the full sample.

In the right column, both simulations show good agreement with the xGASS median for $M_{\star} > 10^{10.3} M_{\odot}$ (i.e., less than 0.1 dex difference), which then increases to a maximum discrepancy at $M_{\star} = 10^{9} M_{\odot}$ (0.35 dex for {\sc TNG} and 0.42 dex for {\sc EAGLE}).
We note that at stellar masses below $\sim 10^{10} M_{\odot}$ the resolution of both simulations has an impact on the AM measurements determined \citep{Wilkinson2023}.
Therefore, the disagreement at low stellar masses should not be over-interpreted, and we conclude that the agreement between xGASS and the simulation data are reasonable for the mock detection sample.

We also show the baryonic Fall relation in Fig. \ref{fig: BaryonicFall}.
We note a better agreement between xGASS and the simulation data for the baryonic Fall relation than the stellar Fall relation.
Without any selection cuts (left column) for {\sc EAGLE}, there is a maximum discrepancy of 0.42 dex at $10^{11} M_{\odot}$, and less than 0.2 dex difference for $M_{\rm{bar}} < 10^{10.6} M_{\odot}$. 
{\sc TNG} has less than 0.2 dex discrepancy at all baryonic masses.
When the mock detection sample is considered (right column), any disparities between either of the simulations and the xGASS observations are effectively eliminated. 
Specifically, {\sc TNG} and {\sc EAGLE} have discrepancies of less than 0.1 dex and less than 0.2 dex at all baryonic masses, respectively.
This better agreement for the baryonic Fall relation is especially important for this work, as the remainder of the analysis will focus on the baryonic j.

\subsection{$j_{\rm{bar}}$--$M_{\rm{bar}}$--$f_{\rm{atm}}$ plane}
\begin{figure*}
    \centering
    \includegraphics[width=\textwidth]{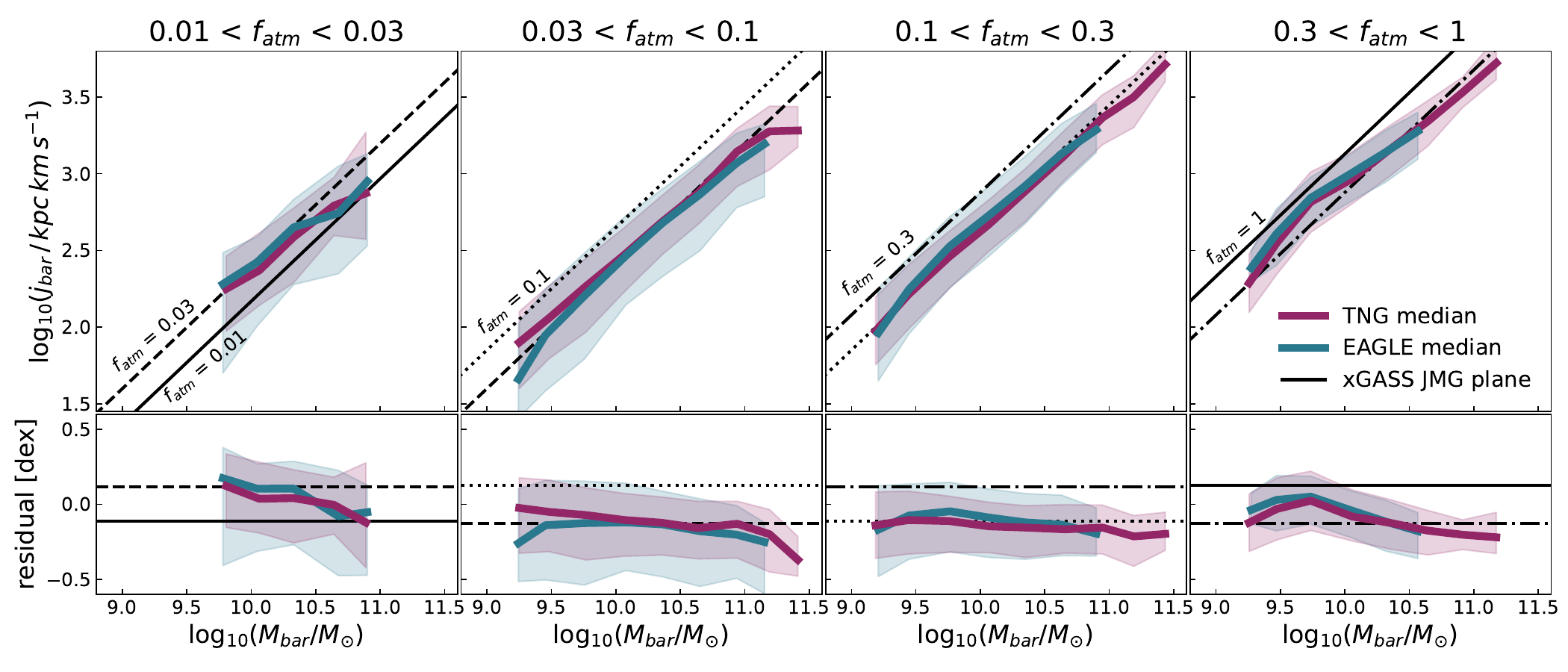}
    \caption{The top row shows sliced projections of the $j_{\rm{bar}}$--$M_{\rm{bar}}$--$f_{\rm{atm}}$ plane, with baryonic mass on the x-axis, baryonic specific AM on the y-axis, and where each column is a range of atomic gas fractions (evenly log-spaced). {\sc TNG} and {\sc EAGLE} medians in bins of 0.3 dex in $M_{\rm{bar}}$ are shown as magenta/ blue lines respectively, with the 16th and 84th percentiles for each of these bins shown as the shaded regions. Only the mock detection sample is shown (see text). For comparison, we show the xGASS JMG plane for fixed gas fractions. This is shown as black lines for the gas fractions  of the bin edges for each column. The bottom row shows the vertical residual (in dex) with respect to the midpoint of the JMG plane for that gas fraction bin.}
    \label{fig: Both_3Dplane}
\end{figure*}

In this subsection, we now compare both {\sc EAGLE} and {\sc TNG} to the $j_{\rm{bar}}$ - $M_{\rm{bar}}$ - $f_{\rm{atm}}$ plane, which was found for xGASS in \citetalias{Hardwick2022b}.
The top row of Fig. \ref{fig: Both_3Dplane} shows $j_{\rm{bar}}$ against $M_{\rm{bar}}$ in four evenly log-spaced $f_{\rm{atm}}$ bins.
Each column shows medians (in 0.3 dex $M_{\rm{bar}}$ bins) of the {\sc EAGLE} and {\sc TNG} galaxies in that atomic gas fraction bin.
The median in each bin is shown by a blue ({\sc EAGLE}) or magenta ({\sc TNG}) line, and the shaded regions show the range of the 16th to 84th percentile for each bin.
For comparison, the black lines show the xGASS JMG plane at fixed gas fractions. 
We show these lines for the gas fractions at the bin edges of each of the columns ($f_{\rm{atm}} =$ 0.01, 0.03, 0.1, 0.3 is solid, dashed, dotted and dot-dashed, respectively).
In the bottom row of Fig. \ref{fig: Both_3Dplane} we show the residual in dex of these medians with respect to the midpoint of the xGASS JMG plane lines.

In Fig. \ref{fig: Both_3Dplane} we only show the mock detection sample, as, in this projection and gas fraction range, there is very little difference between the full simulation sample and the mock detection sample. 
The mock detection sample is simply a cut in \HI\ gas fraction with a stellar mass weighting applied. 
Therefore, the only difference that can be seen between the two samples, is more galaxies in the first column for the full simulation samples that extend to lower baryonic masses.
However, for completeness in appendix \ref{appendix: ExpandedSimPlanes} we also show the spread of the galaxies in this parameter space as 2D histograms for both the full sample and the mock detection sample, ({\sc EAGLE} and {\sc TNG} are shown as Fig. \ref{fig: EAGLE_3Dplane} and \ref{fig: TNG_3Dplane} respectively).

Regardless of whether the full sample or mock detection samples are used, for gas fractions greater than 0.01, the JMG plane is in excellent agreement for both the simulations and xGASS.
The two middle panels of Fig. \ref{fig: Both_3Dplane} show simulation medians with the same slope as the xGASS plane.
The left and right panels have slopes that are slightly shallower than the xGASS plane but are consistent within their scatter.
In the left panel, this slight difference in slope is simply due to low statistics in this $f_{\rm{atm}}$ bin, which can be seen by the large shaded region.
In the right panel, different values of $f_{\rm{atm}}$ dominate at various baryonic masses, leading to a change in slope. 
Specifically, galaxies with $f_{\rm{atm}} \approx 1$ are predominantly within the range of $9.5 < \log_{10}(M_{\rm{bar}}/ M_{\odot})< 10$.
The peak at $\log_{10}(M_{\rm{bar}}/M_{\odot}) \approx 9.8$ is where these galaxies dominate the median and drive up the average $j_{\rm{bar}}$ at this location.
In all panels, the simulation medians have an offset of less than $0.2$ dex from the allowed region of the xGASS JMG plane.

There are small offsets in normalisation, but these are consistent with the offsets observed in gas fraction and $j$ in Fig. \ref{fig: HIfraction-Mass} and \ref{fig: FallRelations}.
In many ways, the agreement between {\sc EAGLE}, {\sc TNG} and xGASS in this parameter space is remarkable. 
Firstly, it is interesting that both simulations are in good agreement with each other (maximum difference of 0.23 dex between the two simulations' medians) given that they rely on different codes and subgrid physics prescriptions. 
Often these differences will result in the simulations having slightly different predictions, such as the \HI\ fraction -- stellar mass relation (Fig. \ref{fig: HIfraction-Mass}), where low stellar mass {\sc EAGLE} galaxies are more gas poor than {\sc TNG} galaxies. 
However, when gas fraction, baryonic mass, and baryonic specific AM are all considered together, as we have in Fig. \ref{fig: Both_3Dplane}, there is very little difference between the two simulations' predictions.
Secondly, it is also intriguing how well these simulations agree with the xGASS JMG plane. 
In particular, this agreement is strongest when considering $j_{\rm{bar}}$, $M_{\rm{bar}}$ and $f_{\rm{atm}}$ together, rather than when isolating either the \HI -- stellar mass relation (Fig. \ref{fig: HIfraction-Mass}) or Fall relation (Fig. \ref{fig: FallRelations}) separately.

It should be emphasised that, despite {\sc EAGLE} and {\sc TNG} having their subgrid models calibrated against many observational scaling relations \citep{Crain2015,Schaye2015,Pillepich2018b,Nelson2018}, this is likely not the cause of the tight agreement seen between the simulations and observations in Fig. \ref{fig: Both_3Dplane}.
First, although both simulations are calibrated to reproduce the observed stellar mass--size relation, which is closely linked to the Fall relation, Fig. \ref{fig: StellarFall} and \ref{fig: BaryonicFall} show that the simulated Fall relations have larger discrepancies between the simulations and observations than is seen for the JMG plane in Fig. \ref{fig: Both_3Dplane}.
Therefore, the simulations calibration to reproduce the stellar mass--size relation is not the sole cause of the tight agreement between the JMG plane and {\sc EAGLE}/{\sc TNG}.
Second, neither of these simulations are calibrated to reproduce cold gas content, as the majority of the subgrid models calibrate for stellar content.
Therefore, the tight agreement seen between simulations and observations in the $j_{\rm{bar}}$, $M_{\rm{bar}}$ and $f_{\rm{atm}}$ plane is not due only to the simulations calibration, and is instead a prediction of the models.

We also chose to fit a JMG plane directly to the simulation data using the \textsc{hyper-fit} \citep{Robotham2015} Bayesian hyperplane fitting tool, as we did for xGASS data in \citetalias{Hardwick2022b}.
As the full simulation data are heavily skewed in their baryonic mass and AM distribution, we choose to only fit a JMG plane to the mock detection sample, (we explore this skewness more in section \ref{section: scatter}).
The best fitting parameters of a JMG plane with the form 
\begin{equation}
    \log_{10}(j_{bar}) = \alpha \log_{10}(M_{bar}) + \beta \log_{10}(f_{atm}) + \gamma
    \label{eq: plane_equation}
\end{equation} 
are shown in table \ref{tab: 3Dplanes}. 
\begin{table*}
    \centering
    \begin{tabular}{ccccc}
        \hline
                     &     $\alpha$    &    $\beta$      &     $\gamma$ & $\sigma$\\
        \hline
        {\sc TNG} & $0.78 \pm 0.01$   & $0.63 \pm 0.02$ & $-4.70 \pm 0.14$ & $0.17 \pm 0.01$  \\
        {\sc EAGLE} & $ 0.94 \pm 0.02$ & $0.90 \pm 0.04$ & $-6.02 \pm 0.24$  & $0.25 \pm 0.01$ \\
        xGASS \citep{Hardwick2022b}  & $0.80 \pm 0.02$ & $0.48 \pm 0.02$ & $-4.86 \pm 0.16$ & $0.15 \pm 0.01$ \\
        \hline
    \end{tabular}
    \caption{The coefficients of the best-fitting JMG plane of the form $\log_{10}(j_{bar}) = \alpha \log_{10}(M_{bar}) + \beta \log_{10}(f_{atm}) + \gamma$ to the data in the left column. $\sigma$ is the standard deviation in the vertical (j) direction.}
    \label{tab: 3Dplanes}
\end{table*}
These values can be compared to those found for \citetalias{Hardwick2022b} in the bottom row of the table.
We see that the {\sc TNG} JMG plane parameters are within errors of the xGASS JMG plane (except for the $\beta$ parameter), while the {\sc EAGLE} simulation parameters deviate by more than $3\sigma$. 
Although, it should be noted that the errors provided on these parameters are uncertainties from the MCMC (Markov Chain Monte Carlo) chain and do not incorporate any uncertainties on individual galaxy values, so will likely be an underestimate of the true error.
The projection of both of these fits are shown in Fig. \ref{fig: EAGLE_3Dplane} and \ref{fig: TNG_3Dplane}, where it is clear that the best fit is not always an accurate representation of the data distribution, in particular at low gas fraction, so that median values provide a more fair comparison. 
As we will see in the next section, this is also because at low gas fractions the JMG plane may no longer be able to properly describe the distribution of galaxies in both EAGLE and TNG.
We can also compare the standard deviation of galaxies from the JMG plane in the $j$ direction, which is given in the right column of table \ref{tab: 3Dplanes}.
The spread of galaxies is the smallest in xGASS, with {\sc TNG} being marginally larger and {\sc EAGLE} being $\sim50$\% larger.
In an attempt to better understand these differences, in the next subsection, we look into the scatter around the JMG plane in more depth.

\subsection{Scatter from the JMG plane} \label{section: scatter}

\begin{figure*}
    \centering
    \includegraphics[width=\textwidth]{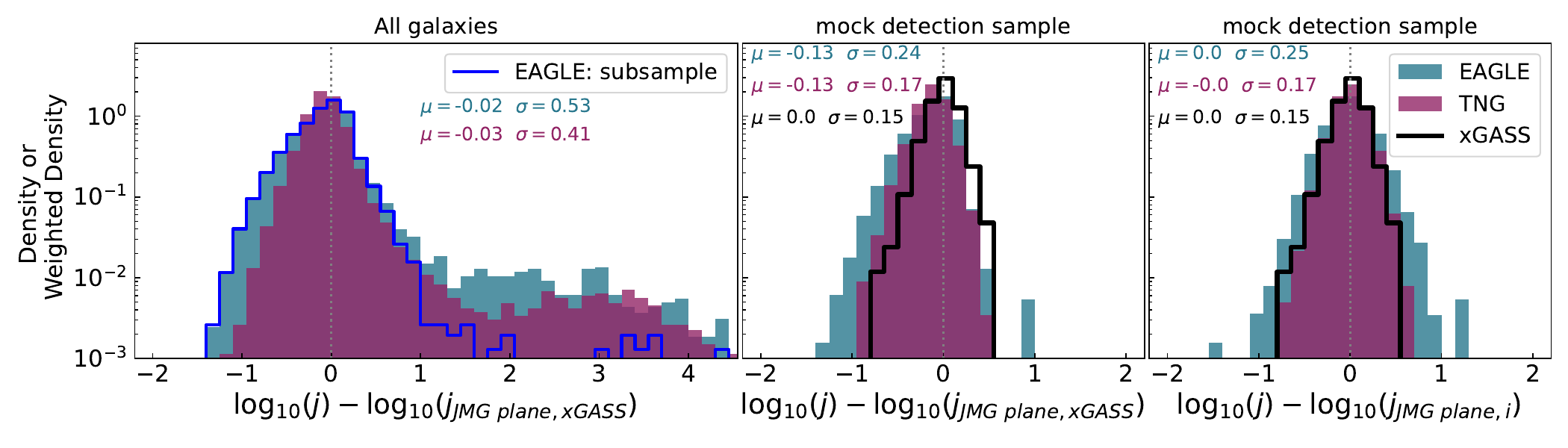}
    \caption{Histograms of offsets in the j direction from the JMG plane. The left and middle columns are offsets from the xGASS JMG plane, while the right column is offset from the JMG plane of their respective simulation. The left column shows all {\sc TNG} and {\sc EAGLE} galaxies as a density histogram, while the middle and right column is weighted histograms for the mock detection samples. In the left column, a blue hollow histogram shows a sub-sample of {\sc EAGLE} that has a reliable gas fraction; i.e., these galaxies contain at least 10 gas particles that have at least 5\% of their mass in neutral gas. In the middle and right columns, we show a comparison to the xGASS sample as the black hollow histogram.}
    \label{fig: OffsetsFrom3DPlane_Both}
\end{figure*}
In addition to studying the shape and slope of the JMG plane, it is also important to investigate the scatter around it.
The offsets of galaxies from the xGASS JMG plane in the $j$ direction are shown in the left and middle panels of Fig. \ref{fig: OffsetsFrom3DPlane_Both}. 
The only difference between the two panels is that the left panel shows all galaxies within either the {\sc EAGLE} or {\sc TNG} simulations (blue and magenta histograms, respectively), whereas the middle panel shows only galaxies that are in the mock detection samples.

When all galaxies in the simulations are considered, the offsets are strongly positively skewed.
This is due to the simulations containing galaxies that have extremely low gas fractions, while still maintaining a moderate baryonic AM.
In fact, the majority of the galaxies in this long extended tail have gas fractions that are at or below the limit of where they would be considered accurate.
One way to determine if a gas fraction is "accurate" is to count the number of gas particles each contributing at least 5 per cent of their mass to neutral gas. 
The blue hollow histogram shows the result of excluding all galaxies with less than 10 gas particles that reach this criterion for EAGLE.
This illustrates that only a small fraction of galaxies in this extended tail have a sufficient number of gas particles to be reliable.
Therefore, when considering the tail of the entire sample, the precise position of a galaxy with respect to the plane should be regarded as an approximation.
We will further explore this result in section \ref{section: discussion}.

Although this extended tail is dominated by galaxies with uncertain gas masses, there is still a small number of galaxies with reliable gas masses and high offsets.
This could indicate that despite gas fraction being a strong predictor of a galaxy's baryonic AM for galaxies with a gas fraction greater than $\sim$0.01, this relationship breaks down for galaxies with lower gas fractions.
In practice, this is unsurprising as, by construction, the dominant mass component will set the baryonic AM of a galaxy. 
In the gas-poor regime, the gas is no longer the dominant mass component, therefore the JMG plane will be inaccurate.
However, the interesting point to note is that not all galaxies in {\sc EAGLE} and {\sc TNG} will be slow rotators once they have depleted their gaseous reservoir.
We will explore this more in section \ref{section: discussion}, but to summarise, this has two implications; first, the JMG plane is only valid for galaxies that have a significant fraction of their mass in \HI\ and second, the physical connection between baryonic AM and gas fraction is not universal for all galaxies.
In summary, the xGASS JMG plane is only applicable to galaxies with a gas fraction greater than $\sim$0.01.

This limit is close to the limit applied to create our mock detection sample.
In the middle panel of Fig. \ref{fig: OffsetsFrom3DPlane_Both}, we show that once galaxies with low gas fractions are removed from the simulation samples, then the long positive skewed tail is removed. 
The offsets now have a distribution that is much closer to a normal distribution but it is not centred on zero.
This can be seen by the values printed in the top left corner, which show the mean ($\mu$) and standard deviation ($\sigma$) of a Gaussian fit to these offsets, (in blue and magenta for {\sc EAGLE} and {\sc TNG} respectively).
This shows that both simulations on average lie $\sim$0.1 dex below the JMG plane.
The standard deviation of these offsets also implies that {\sc EAGLE} has a larger spread around the JMG plane ($\sigma = 0.23$) than {\sc TNG} ($\sigma = 0.18$).

In the right column of Fig. \ref{fig: OffsetsFrom3DPlane_Both}, instead of showing both simulations' offsets from the best-fitting JMG plane to xGASS, we now compare the offset of the simulations from the best fitting JMG plane to their own data.
Although this does not significantly affect the standard deviation of these offsets, they are now centred on zero.

We also compare the offsets of the xGASS galaxies from the xGASS JMG plane in the middle and right panels of Fig. \ref{fig: OffsetsFrom3DPlane_Both} with the black hollow histograms.
The xGASS observational vertical scatter is $\sigma =0.15$. 
This is similar to the spread found for {\sc TNG} ($\sigma = 0.17$) but is smaller than the {\sc EAGLE} spread ($\sigma = 0.23$).
When we calculated the xGASS JMG planes' $\sigma$, we did not attempt to calculate an intrinsic scatter (which takes into account the observational errors of the values), meaning this should instead reflect a combination of measurement errors as well as intrinsic variations of galaxies from the JMG plane.
This was due to us not being confident in the exact values of our errors, as it was difficult to combine observational errors, methodology uncertainty and errors introduced from assumptions for individual galaxies, (see \citetalias{Hardwick2022a} and \citetalias{Hardwick2022b} for more details). 
For the simulations, it is also hard to determine the exact errors on each galaxy's AM, as there is error associated with splitting the gas particles into phases (as we will elaborate on in section \ref{section: discussion}), the uncertainty in the assumptions used to determine AM and particle shot noise.
Therefore, it is unclear if the difference in scatter seen between the observations and simulations is statistically significant.

\subsection{JMG plane offsets and their relationship to star formation rates} \label{section: SFR_vs_scatter}

The next science question that simulations and their increased statistics allow us to explore is; "Is there any residual dependence on SFR that the JMG plane does not encapsulate?"
To address that question we look at the distribution of galaxies with respect to the JMG plane and their $\Delta$MS (offset from the star-forming main sequence). 
We attempted to do this analysis with xGASS in \citetalias{Hardwick2022b} but were limited in our statistics, so couldn't draw any strong conclusions.
Given that these simulations have a factor of 20 higher statistics, this should not be an issue with these simulation data.
\begin{figure}
    \centering
    \includegraphics[width=0.5\textwidth]{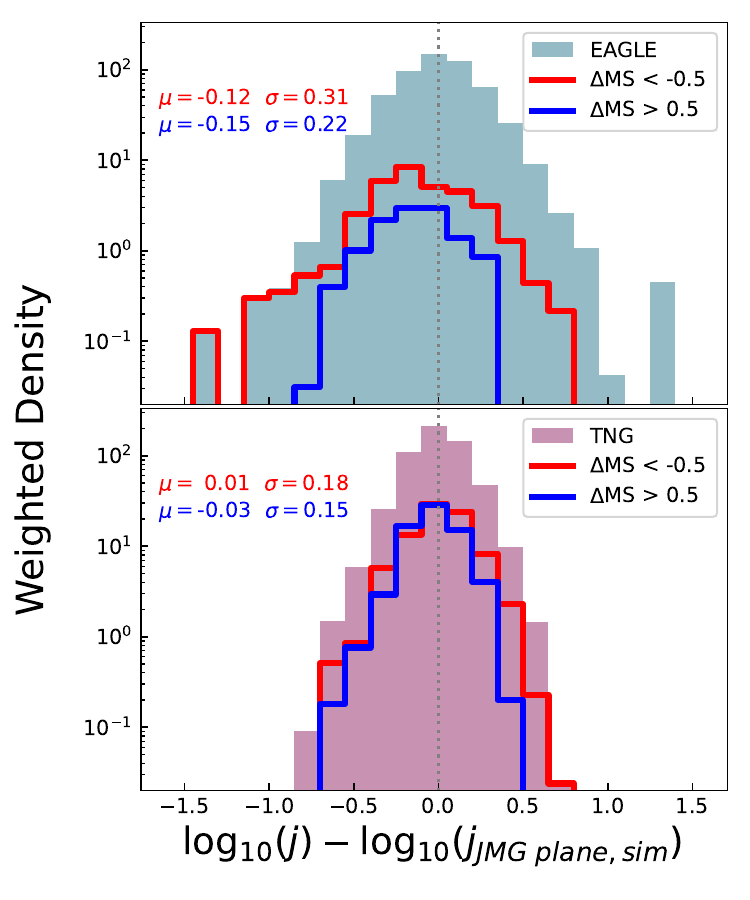}
    \caption{The offset of galaxies from the simulation JMG planes for {\sc EAGLE} (top row) and {\sc TNG} (bottom row). The background histograms are the same as what is shown in the right column of Fig. \ref{fig: OffsetsFrom3DPlane_Both}, but now overlaid we also show the sub-samples that are 0.5 dex above the star-forming main sequence (blue) or 0.5 dex below it (red).}
    \label{fig: Offset_DMS_bins}
\end{figure}

In Fig. \ref{fig: Offset_DMS_bins} we investigate whether variations above and below the star-forming main sequence influence a galaxy's position relative to the JMG plane.
In this figure, we show {\sc EAGLE} (top row) and {\sc TNG} (bottom row) offsets for the mock detection sample (same as those in the right panel of Fig. \ref{fig: OffsetsFrom3DPlane_Both}).
This shows offsets from either the {\sc EAGLE} or {\sc TNG} JMG planes (for the top and bottom rows, respectively).
Overlaid on these background histograms are subsamples of galaxies that are either 0.5 dex above (blue) or below (red) the star-forming main sequence.
The main sequence is re-defined for each simulation's mock detection sample using the curved main sequence from \cite{Leslie2020}.\footnote{This result qualitatively does not change irrespective of whether a linear or curved main sequence is used.} 
In {\sc TNG}, the mock detection sample, above the MS sub-sample and below the MS sub-sample have a similar distribution around the JMG plane.
For {\sc EAGLE}, galaxies below the MS have a slightly larger spread around the plane than galaxies above but have a similar mean.
These result qualitatively does not change if we consider offsets from the JMG plane in the $f_{\rm{atm}}$ direction.
This implies that variations in SFR around the main sequence do not seem to be mirrored by variations around the JMG plane, and vice-versa, and that structure and SFR are not directly influencing each other.

\section{Discussion} \label{section: discussion}

\begin{figure}
    \centering
    \includegraphics[width=0.5\textwidth]{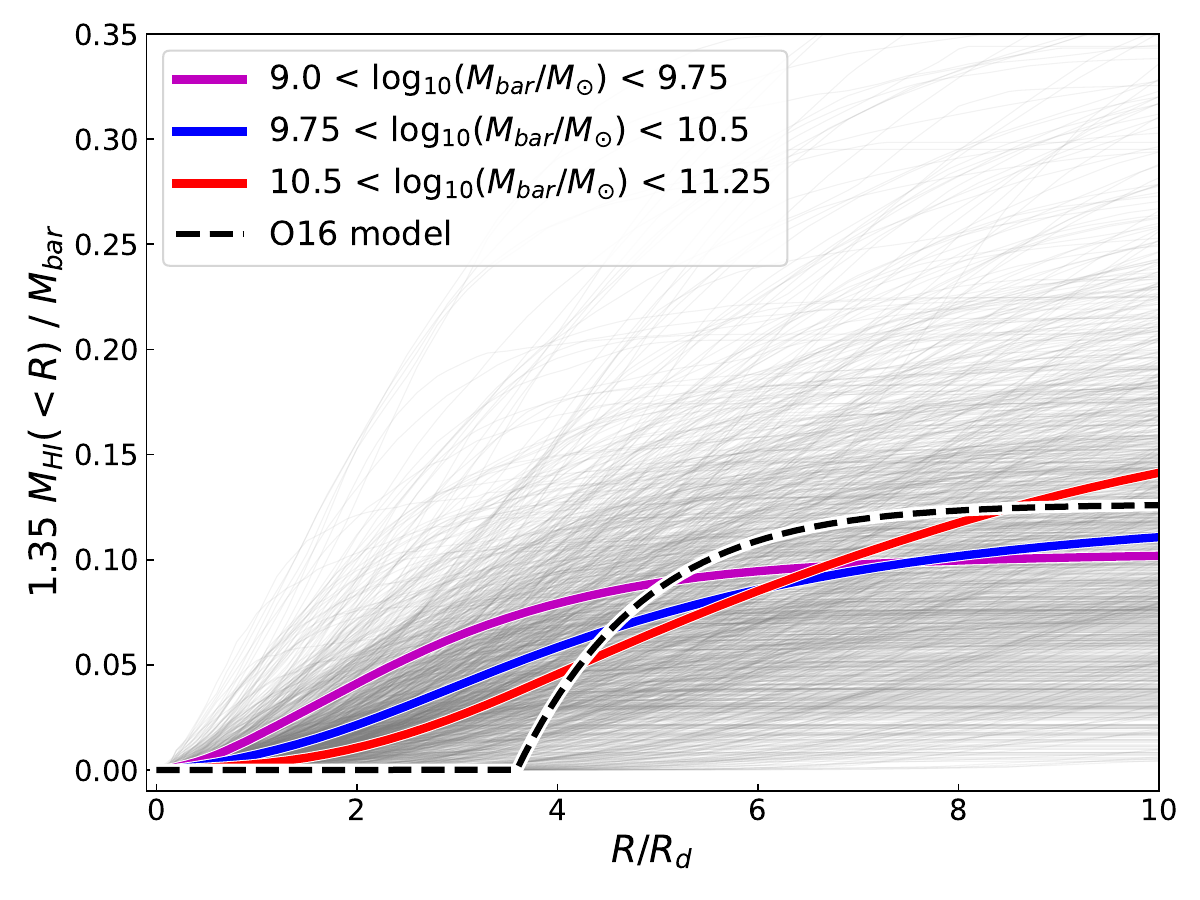}
    \caption{The cumulative atomic gas fraction as a function of radius (normalised by disk radius) of {\sc TNG} galaxies. All of the galaxies that have $0.06 \leq q \leq 0.08$ (N=1445) are shown in grey. The prediction from the \citetalias{Obreschkow2016} model for this bin is shown as the black dashed line. The radially binned median for galaxies within a baryonic mass bin is shown as the coloured lines (see legend for mass bins).}
    \label{fig: HIProfiles}
\end{figure}

Our analysis has shown that in both {\sc EAGLE} and {\sc TNG}, galaxies with gas fractions greater than 0.01 lie on a $M_{\rm{bar}}$ - $j_{\rm{bar}}$ - $f_{\rm{atm}}$ plane that is remarkably similar to the empirical one found in \citetalias{Hardwick2022b}.
Therefore, the tension we found in \citetalias{Hardwick2022b} between the \cite{Obreschkow2016} gravitational stability model (hereafter \citetalias{Obreschkow2016} model) is also true for {\sc EAGLE} and {\sc TNG}.
The \citetalias{Obreschkow2016} model has the same qualitative trends as our simulation data but the exact exponents of the \citetalias{Obreschkow2016} model differ from what we find for {\sc EAGLE} and {\sc TNG}.
As we already discussed in \citetalias{Hardwick2022b}, the \citetalias{Obreschkow2016} model makes several simplifying assumptions for the sake of an analytical argument. 
These assumptions result in the model predicting a single \HI\ profile shape given a fixed $q:= j_{\rm{bar}} \sigma/G M_{\rm{bar}}$.
By using the increased statistics of the simulation data, and its ability to accurately resolve cold gas profiles, we find that the \HI\ profiles have additional baryonic mass dependence which is not encapsulated in this model.
A similar result was found in \citealt{Stevens2018} with halo mass.
An example of {\sc TNG} \HI\ profiles within a small $q$-interval is shown in Fig. \ref{fig: HIProfiles}.
This shows that different baryonic masses exhibit differently shaped \HI\ profiles.
The discrepancy between \citetalias{Obreschkow2016} and the cosmological simulations might be due to this non-trivial scale-dependence.

The agreement between xGASS, {\sc TNG} and {\sc EAGLE} suggested that we can use the simulations to gain a deeper understanding of the properties and shape of the JMG plane and its implications for galaxy evolution.

As we have shown in Fig. \ref{fig: Both_3Dplane}, galaxies in the $M_{\rm{bar}}$--$j_{\rm{bar}}$--$f_{\rm{atm}}$ parameter space are well described by a plane (in log space) in both simulations for gas fractions above 0.01. 
However, below this threshold there is an indication that galaxies may deviate from the JMG plane, as illustrated by the long tail in the distribution shown in the left panel of Fig. \ref{fig: Both_3Dplane}.
Galaxies with very low gas fractions have higher baryonic specific AM than predicted by the JMG plane.
Although, it should be noted that these galaxies have gas fractions that are at the limit of what would be considered reliable (see results section).
Since the simulations do not directly model gas phases, the \HI\ mass of a galaxy is determined in post-processing, with each gas particle assigned a percentage for neutral and then atomic gas.
For galaxies to have such low gas fractions, they have either a small number of gas particles contributing to the atomic mass and/or each gas particle contributes a very small percentage of its mass to atomic gas.
In the very gas--poor regime, potential errors in the gas phase splitting and particle shot noise add significant statistical and systematic uncertainties to the gas fraction. 
To address this, in Fig. \ref{fig: Both_3Dplane} we also show the blue hollow histogram for a sub-sample of galaxies in {\sc EAGLE}, where uncertain \HI\ masses are removed.
In this figure, we define this sub-sample as galaxies that have at least 10 gas particles, each with at least 5\% of their mass in neutral gas. 
This approach provides a more conservative representation of the distribution of {\sc EAGLE} galaxies around the plane, now showing only a faint indication of a tail with large offsets.\footnote{We investigated variations in the percentage of neutral mass required for gas particles to be considered accurate (10\% and 25\%) as well as the number of particles requiring this percentage (20 and 30) and obtained qualitatively similar results.}
Despite the majority of galaxies within this long extended tail having uncertain \HI\ masses, there are still 38 galaxies with offsets greater than 1 dex and reliable gas fractions.
This provides speculative evidence that gas fraction may not accurately predict a galaxy's baryonic AM in the gas--poor regime.
These reliable large offset galaxies all have low gas fractions ($f_{\rm{atm}} < 10^{-3}$), whilst possessing moderate baryonic AM.
Therefore, whatever mechanism caused the galaxies to deplete their gaseous reservoirs did not significantly reduce their baryonic AM.
We inspected these outlier galaxies, and they all appear to be quiescent disc galaxies with the majority of their velocity fields showing regular rotation.
Only 2 have undergone a merger (i.e., stellar mass ratio greater than 0.1) in the last gigayear.

\begin{figure*}
    \centering
    \includegraphics[width=\textwidth]{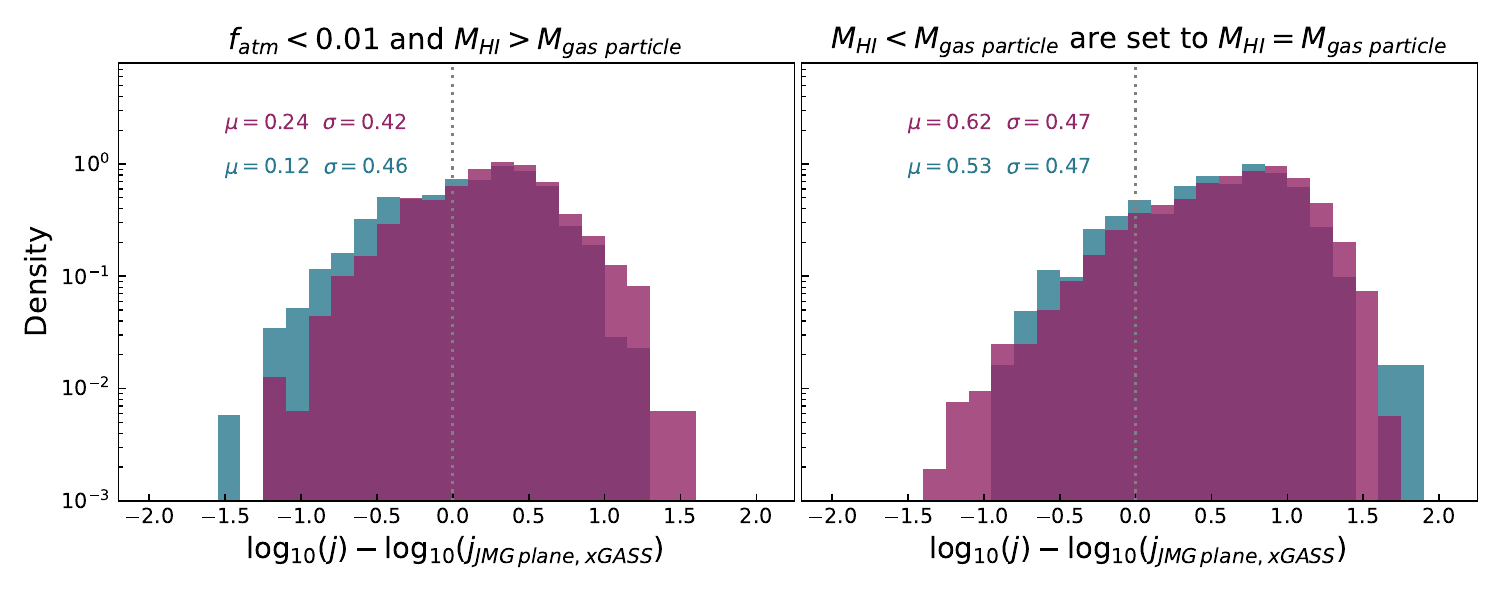}
    \caption{The offset of gas--poor galaxies (i.e., $f_{atm} < 0.01$) from the JMG plane, for {\sc EAGLE} (blue) and {\sc TNG} (magenta). The left panel shows all the galaxies which have \HI\ mass greater than the mass of one gas particle. In the right column, these are galaxies with uncertain \HI\ masses (i.e., $M_{\rm{HI}} < M_{\rm{gas\ particle}}$). For these galaxies, we set the \HI\ mass to be an upper limit of $M_{\rm{HI}} = M_{\rm{gas\ particle}}$. The mean ($\mu$) and standard deviation ($\sigma$) of these distributions are printed in the top left corner.}
    \label{fig: UpperLimits}
\end{figure*}

It is worth noting that although the gas fractions of most gas-poor galaxies are uncertain (and consequently, their exact offsets from the JMG plane are uncertain), the galaxies as a whole are still resolved, making it unlikely for them to become gas--rich if a higher resolution simulation was conducted.
In addition, as their $j_{\rm{bar}}$ and $M_{\rm{bar}}$ are dominated by stars, these values can be considered accurate.
We find that gas--poor galaxies maintain a similar $j_{\rm{bar}}$ -- $M_{\rm{bar}}$ distribution, once they fall below a gas fraction of 0.01.
Therefore, excluding all of these gas--poor galaxies from our analysis limits the conclusions that we can draw and is somewhat unnecessary.
To address this, in Fig. \ref{fig: UpperLimits}, we adopt a slightly different reliability measure; we assume that all galaxies with a total \HI\ mass greater than the mass of one gas particle to be accurate (for {\sc EAGLE} $M_{\rm{gas\ particle}} = 1.81 \times 10^{6} \rm{M}_{\odot}$ and for {\sc TNG} $M_{\rm{gas\ particle}} = 1.4 \times 10^{6} \rm{M}_{\odot}$).
Galaxies with \HI\ masses below this threshold are assigned an upper limit of $M_{HI} = M_{\rm{gas\ particle}}$.
This alternative criterion obtains a similar result to setting a limit based on the number of particles, despite not explicitly checking for an adequate number of gas particles, and allows us to easily apply the same condition to both {\sc EAGLE} and {\sc TNG}.
Fig. \ref{fig: UpperLimits} shows galaxies with $f_{\rm{atm}} > 0.01$, with the left panel showing all galaxies with $M_{\rm{HI}} > M_{\rm{gas\ particle}}$ and the right panel the galaxies with \HI\ masses set to the upper limit.
This figure shows that the distribution of $f_{\rm{atm}} < 0.01$ galaxies is preferentially above the JMG plane. 
$M_{\rm{HI}} > M_{\rm{gas\ particle}}$ galaxies are, on average, 0.12 and 0.24 dex above the JMG plane, while galaxies set to the upper limits are 0.53 and 0.62 dex above the plane (for EAGLE and TNG, respectively).
In other words, galaxies with low gas fractions possess higher AM at a fixed mass and gas fraction than expected for that gas fraction.
We expect that the true distribution of gas--poor galaxies around the plane would be somewhere in between the distributions seen in Fig. \ref{fig: OffsetsFrom3DPlane_Both} and Fig. \ref{fig: UpperLimits}.
Once upcoming cosmological simulations that model gas phases directly become available, we will be able to determine the exact relationship between AM and gas content in the gas--poor regime.

This result is noteworthy, as we see that a considerable number of galaxies have moderate angular momentum values despite having little-to-no gas.
If all gas-poor galaxies were slow rotators, the JMG plane would still hold in the gas-poor regime, which is not the case.
Therefore, despite gas being a strong indicator of a galaxy's angular momentum in the gas-normal regime, it breaks down in the gas-poor regime.
A similar result was found in \citetalias{Hardwick2022a} where we showed that the scatter of the stellar Fall relation was strongly correlated with \HI\ gas fraction at low masses ($M_{\star} < 10^{10.25} M_{\odot}$), even more so than bulge-to-total ratio.
However, when we looked at the high mass regime  ($M_{\star} > 10^{10.25} M_{\odot}$) then \HI\ gas fraction became less dominant, with bulge-to-total ratio having a slightly stronger correlation with scatter than gas fraction.
  
It is not surprising that both {\sc EAGLE} and {\sc TNG} indicate the presence of a substantial population of passive galaxies with minimal gas but significant angular momentum.
Seminal studies of the Virgo cluster already showed that a large fraction of passive galaxies are structurally more similar to discs than ellipticals \citep{Binggeli1988, Lisker2006}. 
More recently, the advent of integral field spectroscopic (IFS) surveys has firmly established that the vast majority of passive galaxies show stellar angular momenta not too dissimilar from those observed in star-forming galaxies \citep[e.g.,][]{WangB2020, Fraser-McKelvie2022, Cortese2019, Cortese2022}. 
This clearly highlights how star formation quenching and major structural transformation are two separate (and not always associated) processes in the evolution of galaxies.

The fact that the JMG plane is not universal and valid only for a sub-sample of the local galaxy population does not reduce its importance for galaxy evolution studies. 
As both {\sc EAGLE} and {\sc TNG} implement their star formation and feedback processes differently, the fact that both simulations agree so well with each other could imply that the JMG plane is primarily set by gravitational processes and how cold gas settles in galaxies and reaches equilibrium, rather than any major connection with the way \HI\ is used for star formation.

The idea that the shape of the JMG plane is disconnected from the gas--star formation cycle in galaxies (i.e., both quenching and star-forming stages) is further supported by the lack of any correlation between a galaxy's offset from the $j_{\rm{bar}}$ - $M_{\rm{bar}}$ - $f_{\rm{atm}}$ plane and its position with respect to the star-forming main sequence.
In other words, galaxies that have higher SFR with respect to the main sequence, are not preferentially above or below the JMG plane and vice versa.
The trends shown in Fig. \ref{fig: Offset_DMS_bins} are qualitatively the same if offsets are calculated in the $f_{\rm{atm}}$ direction. 
This implies that the physical processes causing an increase (or decrease) in SFR are not driven by processes that cause an increase (or decrease) in atomic gas fraction with respect to the $j_{\rm{bar}}$ - $M_{\rm{bar}}$ - $f_{\rm{atm}}$ plane.
This is interesting given that $\Delta$MS is strongly correlated with offsets from the $M_{\star}$ - $f_{\rm{atm}}$ relation, \citep[e.g.,][]{Saintonge2022} and we see a similar correlation for the offsets from the $M_{\rm{bar}}$ - $f_{\rm{atm}}$ relation for both {\sc EAGLE} and {\sc TNG}.
We can speculate on potential scenarios that could cause the excess cold gas (with respect to the JMG plane) to not be correlated with SFR, such as a large ring of stable \HI\ in the outskirts of a galaxy, which could result in an increase in \HI\ gas fraction, without triggering a starburst event.
However, more work is needed to determine the process (or processes) driving the scatter of this JMG plane, and why it is disconnected from galaxies star formation rates.

\section{Conclusions} \label{section: conclusion}

This study presents a comprehensive comparison between the $j_{\rm{bar}}$ - $M_{\rm{bar}}$ - $f_{\rm{atm}}$ plane for xGASS data presented in \citetalias{Hardwick2022b}, and cosmological simulation data from {\sc EAGLE} and {\sc TNG}. 
We compared all the galaxies in each simulation volume, and mock detection samples, to determine how sample selection could be affecting our results. 
We summarise our main conclusions as follows:
\begin{enumerate}
    \item The $j_{\rm{bar}}$ - $M_{\rm{bar}}$ - $f_{\rm{atm}}$ plane found for the xGASS sample, is consistent in both orientation and scatter with the {\sc EAGLE} and {\sc TNG} mock detection samples and full simulation samples, for $f_{\rm{atm}} > 0.01$.
    \item There is moderate evidence that for gas fractions below $f_{\rm{atm}} \sim 0.01$, the simulations deviate from the empirical JMG plane, asymptoting towards a constant $j_{\rm{bar}}$ - $M_{\rm{bar}}$ distribution that no longer depends on $f_{\rm{atm}}$.
    \item The scatter in this JMG plane is independent of $\Delta$MS (for $f_{\rm{atm}} > 0.01$), suggesting that the processes causing deviations from the star-forming main sequence do not affect the processes causing deviations from the JMG plane.
\end{enumerate}

It would be interesting for future works to investigate tracking these simulated galaxies through different redshift snapshots to see if this gives insights into the drivers of scatter in this JMG plane and the factors contributing to its deviation at low gas fractions.

\section*{Acknowledgements}

We thank the anonymous referee for their comments which improved the quality of this manuscript.
JAH and LC acknowledge support from the Australian Research Council (FT180100066). Parts of this research were conducted by the Australian Research Council Centre of Excellence for All Sky Astrophysics in 3 Dimensions (ASTRO 3D), through project number CE170100013. DO is a recipient of an Australian Research Council Future Fellowship (FT190100083), funded by the Australian Government.
ARHS is funded through the Jim Buckee Fellowship at UWA.
We acknowledge the Virgo Consortium for making their simulation data available. The {\sc EAGLE} simulations were performed using the DiRAC-2 facility at Durham, managed by the ICC, and the PRACE facility Curie based in France at TGCC, CEA, Bruyeres-le-Chatel.
The IllustrisTNG simulations were undertaken with compute time awarded by the Gauss Centre for Supercomputing (GCS) under GCS Large-Scale Projects GCS-ILLU and GCS-DWAR on the GCS share of the supercomputer Hazel Hen at the High Performance Computing Center Stuttgart (HLRS), as well as on the machines of the Max Planck Computing and Data Facility (MPCDF) in Garching, Germany.

\section*{Data Availability}

All of the xGASS data used in this work are publicly available at \url{www.xgass.icrar.org}.
The {\sc EAGLE} simulations are publicly available; see \citet{McAlpine15,EAGLE17} for how to access {\sc EAGLE} data.
For access to the {\sc TNG} data used here, please contact ARHS. Otherwise, the public-facing {\sc TNG} database has similar -- although, not identically calculated -- galaxy properties available at \url{www.tng-project.org/data/}



\bibliographystyle{mnras}
\bibliography{references.bib}



\appendix

\section{Detailed investigation of the simulation planes} \label{appendix: ExpandedSimPlanes} 

\begin{figure*}
    \centering
    \includegraphics[width=\textwidth]{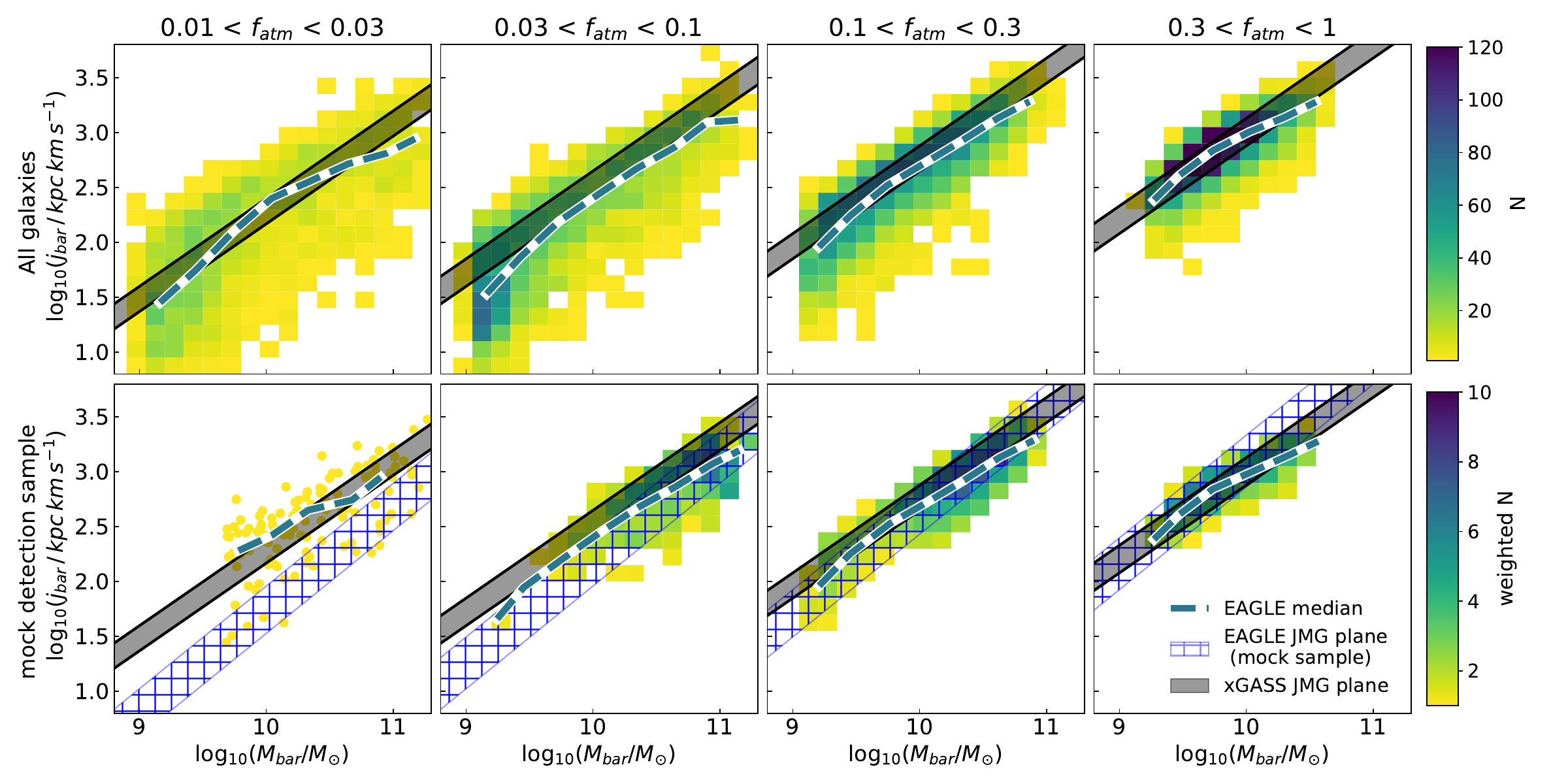}
    \caption{The $M_{\rm{bar}}$ - $j_{\rm{bar}}$ - $f_{\rm{atm}}$ JMG plane for {\sc EAGLE} galaxies. The top row is the full sample in {\sc EAGLE} (i.e., all galaxies with a stellar mass greater than $10^{9} M_{\odot}$). In contrast, the bottom row is only the mock detection sample (see section \ref{section: mock obs sample} for sample selection description). The background shows a 2D histogram (or weighted 2D histogram for the bottom row) showing the distribution of all the galaxies in that panel. If there are less than 200 galaxies in a panel, then the location of the galaxies is shown as yellow points, rather than a 2D histogram. The blue dashed lines show the median in bins of 0.3 dex in $M_{bar}$. The xGASS JMG plane, for the gas fractions range of each panel, is shown as the grey-shaded region. For comparison in the bottom row, the blue hashed region shows the best-fit JMG plane to the {\sc EAGLE} mock detection sample.}
    \label{fig: EAGLE_3Dplane}
\end{figure*}

\begin{figure*}
    \centering
    \includegraphics[width=\textwidth]{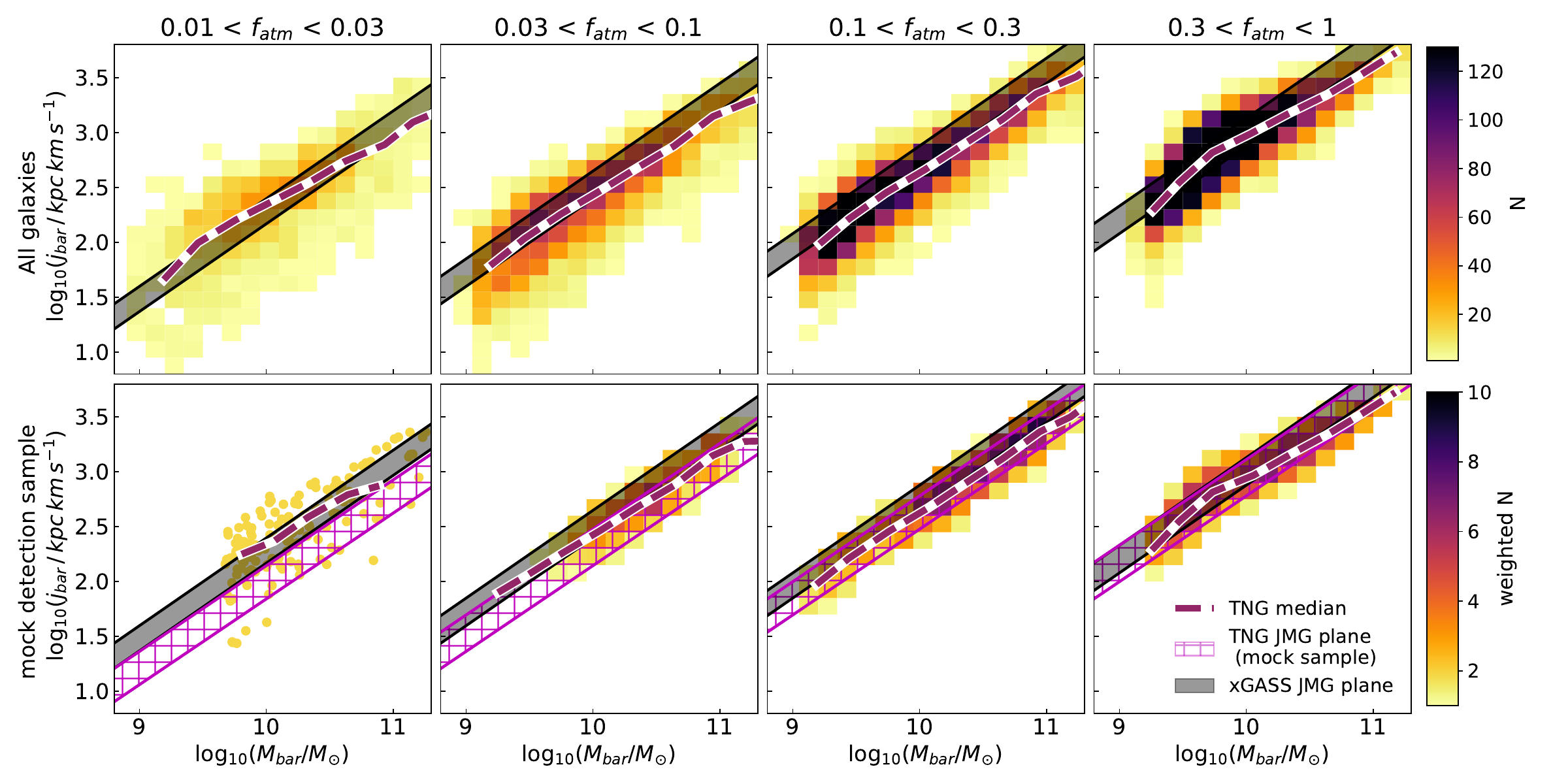}
    \caption{The same as Fig. \ref{fig: EAGLE_3Dplane} but showing {\sc TNG} data and a change of colour scheme. The median distribution and best-fit JMG plane are shown by magenta dashed lines and magenta hashed regions respectively.}
    \label{fig: TNG_3Dplane}
\end{figure*}

For conciseness, in Fig. \ref{fig: Both_3Dplane} we present only the comparison between the xGASS JMG plane and simulations for the binned medians of the mock detection sample.
For completeness, in this section, we also show the full {\sc EAGLE} and {\sc TNG} samples, as well as showing the 2D histogram distribution of the galaxies. 
This is shown in Fig. \ref{fig: EAGLE_3Dplane} and \ref{fig: TNG_3Dplane} for {\sc EAGLE} and {\sc TNG} respectively.

We chose to only show the mock detection sample in Fig. \ref{fig: Both_3Dplane} because, in the gas fraction range that is greater than 0.03, the binned medians of the full sample and the mock detection sample are almost identical, which can be seen when comparing the top and bottom rows of Fig. \ref{fig: EAGLE_3Dplane} and \ref{fig: TNG_3Dplane}.
The only differences are seen for $0.01 \leq f_{atm} \leq 0.03$, where there is a smaller range in $M_{\rm{bar}}$ for the mock detection sample. 
However, the qualitative agreement with the JMG plane is similar for both the mock and full samples.

Fig. \ref{fig: EAGLE_3Dplane} and \ref{fig: TNG_3Dplane} also highlight that the {\sc EAGLE} simulation has a much larger spread in $j_{bar}$ values (at fixed baryonic mass and atomic gas fraction) than {\sc TNG}.
This is most evident when comparing the full samples of both simulations, but is also seen when comparing the mock detection samples.
This large spread in the data for the full simulation samples, and in particular, the non-Gaussianity of the scatter, which, when combined with the asymptotic behaviour of galaxies approaching a fixed $j_{\rm{bar}}$ at extremely low gas fractions, meant that we could not fit a JMG plane directly to the full simulation data.
\textsc{hyper-fit} \citep{Robotham2015} is a Bayesian fitting tool that assumes data to be normally distributed around the JMG plane. 
Although the code will give a mathematically correct solution even when the data is not normally distributed, this solution is not physically meaningful. 
Therefore, we do not show the fits to the full simulation data in this work.
As the mock detection sample is closer to a Gaussian distribution, we fit a JMG plane to these samples.
The parameters of this fit are given in table \ref{tab: 3Dplanes} and shown in the bottom rows of Fig. \ref{fig: EAGLE_3Dplane} and \ref{fig: TNG_3Dplane} as blue and magenta hashed regions (for {\sc EAGLE} and {\sc TNG} respectively).


\bsp	
\label{lastpage}
\end{document}